# Coherent canted ferrimagnetism and higher-order anisotropy in the nodal-line magnetic semiconductor $Mn_3Si_2Te_6$

Chang-woo Cho[1,2,3,+,*], Beomtak Kang[3,*], Ildo Choi[1], Jitae Gwak[1], Jisung Lee[4], Seung-Young Park[4], Seyoung Kwon[5], Sungkyun Park[5], Joonyoung Choi[2,6], Younjung Jo[6], Benjamin A. Piot[7], and Jun Sung Kim[3,+]

[1]*Department of Physics, Chungnam National University, Daejeon 34134, Korea*
[2]*Institute for Sciences of the Universe, Chungnam National University, Daejeon 34134, Korea*
[3]*Department of Physics, Pohang University of Science and Technology, Pohang 37673, Korea*
[4]*Center for Scientific Instrumentation, Korea Basic Science Institute, Daejeon 34133, Korea*
[5]*Department of Physics, Pusan National University, Busan 46241, Korea*
[6]*Department of Physics, Kyungpook National University, Daegu 41566, Korea*
[7]*Laboratoire National des Champs Magnétiques Intenses, CNRS, LNCMI, Université Grenoble Alpes, Univ Toulouse 3, INSA Toulouse, EMFL, F-38042 Grenoble, France*

## Abstract

The interplay between magnetic order and electronic topology in van der Waals materials enables extreme responses to external stimuli. The nodal-line semiconductor $Mn_3Si_2Te_6$ exemplifies this, exhibiting colossal angular magnetoresistance (CAMR) where resistivity changes by orders of magnitude upon rotating the magnetic field. While this phenomenon implies a profound coupling between spin orientation and charge transport, the microscopic magnetic potentials driving spin orientations remain elusive. Here, we combine thermodynamic torque magnetometry and electron spin resonance spectroscopy to reconstruct the magnetic anisotropy energy that controls magnetization rotation in $Mn_3Si_2Te_6$. We show that low-temperature ground state is a coherent canted ferrimagnet stabilized by competing second- ($K_1$) and fourth-order ($K_2$) magnetic anisotropy. Crucially, torque requires a substantial symmetry-allowed sixth-order term ($K_3$), which provides near-plane stiffness and sustains canting at high fields. Using the resulting anisotropy parameters, we compute the non-linear relation between field angle $\theta_H$ and magnetization angle $\theta_M$ and reparameterize CAMR in terms of $\theta_M$, providing a concrete magnetic basis for how sharp angular transport features can emerge near the in-plane configuration.

---

[+] corresponding author: chang-woo.cho@cnu.ac.kr, js.kim@postech.ac.kr
[*] These authors contributed equally: Chang-woo Cho, Beomtak Kang

**Introduction**

Van der Waals (vdW)-like magnetic semiconductors offer a versatile platform where reduced dimensionality, spin-orbit coupling (SOC), and electronic topology can cooperate to generate unusually anisotropic magneto-transport responses [1-6]. Among them, $Mn_3Si_2Te_6$ has drawn significant attention as a semiconducting ferrimagnet with topological nodal-line features, in which rotating the magnetic field relative to the crystal axes can drive an enormous change in resistivity – referred to as colossal angular magnetoresistance (CAMR) [7-10]. This phenomenology highlights that the magnetization direction can act as a key control parameter for electronic structure and scattering. A central unresolved issue, however, is how the magnetization actually reorients under tilted fields in the experimentally relevant field and temperature ranges [9, 10].

In general, magnetization reorientation is governed by the magnetocrystalline anisotropy, which shapes the angular dependence of the magnetic free energy and sets the equilibrium magnetization direction under an applied field [11]. While a single uniaxial term often yields smooth field tracking, vdW-like magnets frequently host competing anisotropy contributions arising from multi-sublattice magnetism and SOC-driven higher-order terms [12-15]. Such competition can reshape the free-energy profile, reduce the near-hard-axis angular stiffness over certain ranges, and stabilize non-collinear equilibrium states. In this situation, angular transport can become extremely sensitive because the relevant variable for transport is the equilibrium magnetization angle, not simply the applied field angle [16-20].

Previous studies on $Mn_3Si_2Te_6$ have established important pieces of this problem. Angle-dependent transport measurements combined with band-structure calculations have shown that rotating the magnetization can strongly modulate the electronic structure (including SOC-driven reconstructions of magnetic nodal-line bands), providing a natural route to CAMR through magnetization-orientation control [9]. Neutron scattering, on the one hand, has established that the low-temperature magnetic ground state is not a simple collinear ferrimagnet, but exhibits a finite canting of the spins, and has emphasized the role of spin fluctuations and low carrier density in the CAMR regime [21]. Taken together, these works emphasize $Mn_3Si_2Te_6$ as an exceptionally angle-sensitive with a non-collinear spin ground state. Yet, the quantitative form and temperature evolution of the anisotropy energy that governs field-driven magnetization reorientation in the experimentally relevant field window remain unresolved.

Here we combine angular torque magnetometry and low-energy electron spin resonance (ESR) to constrain the anisotropy energy governing reorientation. Torque tracks the global evolution of the magnetization during field rotation through directly observable switching markers, while ESR constrains the local curvature of the free energy near equilibrium. By integrating these complementary thermodynamic constraints, we identify robust signatures of coherent canting, map its evolution with field and temperature, and show that a symmetry-allowed higher-order term is required to reproduce the persistence of switching angles up to 9

T. We further show that this reconstructed magnetization-rotation framework provides a direct, testable connection to the CAMR observed in transport measurements [9].

## Results

**Figure 1a** summarizes the torque configuration and angle definitions. Single crystals (typical lateral size is about 60 um) were mounted on piezoresistive cantilevers such that the external field $H$ rotates in a plane containing the crystallographic $c$-axis and an in-plane direction. We define $\theta_H$ as the field angle measured from the $c$-axis and $\theta_M$ as the equilibrium magnetization angle. **Figure 1b** shows field angle dependence of $\tau/H$ at $H = 0.8$ T for Sample#1 from 2 to 90 K. At 2 K, the angular response is large and markedly non-sinusoidal, with abrupt changes near the hard-axis orientations ($H//c$, $\theta_H = 0º, 180º, 360º$). As temperature approaches $T_c$, the amplitude decreases and the response becomes smoother. This behavior is consistent with the weakening of ferrimagnetic order near $T_c$ that is further supported by the susceptibility as shown in **Fig. 1d**. To expose subtle angular anomalies, **Fig. 1c** presents the angular derivative $d(\tau/H)/d\theta_H$ in polar representations. Above 20 K, the signal is predominantly two-fold symmetric, as expected for a simple easy-plane anisotropy. In contrast, at lower temperature (most clearly at 2 K), the angular positions of extrema shift away from the high-symmetry directions ($\theta_H = 90º, 270º$). In a strictly collinear ferrimagnet with a single uniaxial anisotropy, these extrema would remain pinned to the crystallographic axes. The observed offset therefore provides a purely torque-based indication that the equilibrium magnetization does not coincide with a principal axis even at low fields, which is consistent with a non-collinear canted state reported by neutron scattering study [21].

To track the evolution of the canting-related anomaly, we analyze the angular derivative $d(\tau/H)/d\theta_H$ as a function of $\theta_H$ for multiple fields/temperatures. **Figure 2** presents polar plots of $d(\tau/H)/d\theta_H$ for Sample#2 at selected temperatures between 2 and 85 K, with $H$ spanning 0.5 to 9 T. At 2 K, the derivative exhibits a pronounced two-lobe structure whose characteristic peaks are shifted away from the nominal easy-plane directions as observed in **Fig. 1c**. With increasing field, these shifted features remain robust and evolve systematically rather than collapsing quickly into the high-symmetry angles, indicating that the field-driven approach of saturation magnetization $M$ toward the plane is not trivially described by a single, stiff easy-plane anisotropy. As temperature increases (from 5 to 50 K), the overall amplitude shrinks and the anomaly becomes less distinct, and near the transition (refer to 70 K panel) the response is nearly suppressed. Above the transition (refer to 85 K panel), we do not see such an anomaly anymore, and the tendency of the magnitude of $d(\tau/H)/d\theta_H$ is even reversed. Thus, the torque derivative directly maps the field-temperature evolution of the canted equilibrium state in a way that is difficult to reconcile with a purely collinear easy-plane picture.

For a concise summary of the high-field torque phenomenology, we define an operational switching angle $\theta_c$ directly from the extrema positions in $d(\tau/H)/d\theta_H$ near the in-plane region. **Figure 2b-d** show representative traces at selected temperatures and fields. At each field, two prominent extrema are observed near the in-plane orientation. We denote their positions as $\theta_{c1}$ and $\theta_{c2}$, and use $\theta_c = (\theta_{c2} - \theta_{c1})/2$ as a convenient metric that quantified the separation of the switching markers. This operational definition is robust because it is tied to directly observable features and is rarely affected by the hysteresis between sweep directions above 0.5 T (see **Figure S1** in Supplementary Information). **Figure 2e** compiles $\theta_c(T, H)$ into a compact map that will later be used as the primary high-field constraint when determining higher-order anisotropy.

We next turn to ESR results, which provide an independent, dynamical probe of magnetic anisotropy through the curvature of the free energy at the equilibrium orientation. **Figure 3a-c** show field-frequency color maps of the ESR for Sample#3 at $T = 5$ K measured at three field angles: $H//ab$ ($\theta_H = 90°$), a slightly tilted angle $H//ab$ ($\theta_H = 80°$), and $H//c$ ($\theta_H = 0°$). The color scale represents the relative phase change of a resistive bolometer in a transmission geometry, $\Delta\Phi/\Phi_0$, and the red dashed lines trace the fitted resonance dispersion. In a collinear uniaxial easy-plane ferrimagnet described by a single anisotropy scale, one expects qualitatively distinct Kittel-like behaviors for easy-plane and hard-axis geometries [23]. Specifically, to leading order one expects a square-root-like dispersion for the easy-plane configuration, $\omega_{easy} \propto \sqrt{H(H + M_{eff})}$, where the hard-axis configuration produces a near-linear dependence, $\omega_{hard} \propto (H - M_{eff})$, reflecting the different free-energy curvature for magnetization oscillations about the principal axes [23]. Our data show a strikingly different trend. All three orientations exhibit a concave-down, nearly square-root-like dispersion within the low-field window (up to 1.5 T). In particular, the $H//c$ data in **Fig. 3c** do not follow the near-linear hard-axis behavior expected for a strictly collinear easy-plane state. Instead, we observe the dispersion near zero field closely resembles that for $H//ab$ as presented in **Fig. 3a** and **3b**, implying that the relevant curvature of the free energy around the equilibrium magnetization direction is comparable for these nominally distinct field orientations.

These dispersions indicate that the equilibrium magnetization is already tilted by a finite canting angle $\theta_c$, so that the local free-energy curvature probed by ESR can remain comparable even for nominally distinct field orientations. To parameterize this minimal scenario, we use the phenomenological free energy $F$,

$$F = K_1 \sin^2\theta_M + K_2 \sin^4\theta_M + K_3 \sin^6\theta_M - MH\cos(\theta_H - \theta_M).$$

For each $\theta_H$, the equilibrium orientation $\theta_M$ is obtained from $\partial F/\partial\theta_M = 0$, and the resonance frequency is calculated from the Smit-Beljers formalism using the local curvature of $F$ at equilibrium [24]. Within our ESR window (up to 1.5 T), the resonance curvature has limited sensitivity to the sixth-order term, consequently $K_3$ is poorly constrained by ESR and becomes strongly covariant with $K_2$. We therefore use ESR primarily to constrain the leading anisotropy fields ($K_1/M$, $K_2/M$), while determining $K_3/M$ from torque features in the higher-field and broad-temperature analysis. In practice, we initially fixed $K_3$ to zero and carried out an ESR-only fit (red dashed-lines in **Fig. 3a-c**). Afterwards, to obtain the fitting parameters, we perform a global fit to the 5 K data, following the procedure described in Ref. [22] and Supplementary Information, Section 4. A single global fit to the 5 K maps at $\theta_H = 90°$, $80°$, and $0°$ yields $K_1/M$ = -2.40 T, $K_2/M$ = 1.40 T, and $g_a$ = 2.01. For $H//c$, the same single-mode treatment returns an effective $g_c$ = 0.68, smaller than the reported one [25]. We emphasize, however, that this $g$-factor discrepancy does not affect the main conclusion drawn here. The concave-down $H//c$ dispersion itself is inconsistent with a strictly collinear hard-axis response, supporting finite canting, and the requirement of a symmetry-allowed sixth-order stiffness term is established independently by torque switching markers up to 9 T, as discussed below. We therefore treat $g_c$ as an effective parameter that may absorb multi-sublattice mode mixing, demagnetization/misalignment contributions, and fit covariance within the limited low-field spectral window, and we focus on the anisotropy hierarchy constrained jointly by ESR and torque.

## Discussion

Using the ESR-derived $K_1/M$ and $K_2/M$ values as fixed inputs, we test whether a $K_1$-$K_2$ only free energy can reproduce the observed $\theta_c(T, H)$ over the full field/temperature range. In **Fig. 3d**, we extract the $\theta_c(T, H)$ for selected 2 T, 5 T, 9 T variations from **Fig. 2e**, and compare $\theta_c$ estimated as zero-field limit originated from the ESR fitting parameters. While $K_1$-$K_2$ model captures the presence of canting and its low-field evolution, it typically predicts that $\theta_c$ collapses too rapidly with increasing temperature, particularly when the Zeeman energy approaches the quartic near-plane anisotropy scale. In contrast, the torque switching markers remain clearly separated even at the highest measured fields in our experimental limit (up to 9 T), implying that additional angular stiffness becomes relevant as $\theta_M$ approaches the in-plane configuration ($\sin \theta_M \to 1$). This is precisely the regime where a higher-order term, although nominally small in a low-order expansion sense, can become decisive.

The simplest symmetry-allowed extension is therefore a sixth-order term $K_3 \sin^6 \theta_M$, which selectively reshapes the free-energy curvature in the vicinity of in-plane ($\theta_M \approx 90°$) without strongly altering the lower-angle behavior that ESR constraints. This explains why ESR pins down $K_1/M$ and $K_2/M$ reliably, while torque switching has strong leverage on $K_3/M$. Operationally, we fix $K_1/M$ and $K_2/M$ to the ESR constraints and determine an effective field-independent $K_3/M$ by minimizing the least-square residual between the forward-model and measured $\theta_c(H)$ at 5 K, 20 K, and 40 K, respectively (see Supplementary Information, Section 5). The resulting $K_1/M$, $K_2/M$, and $K_3/M$ are presented in **Fig. 3e**. It is worthwhile to mention that although $K_3$ is a higher-order coefficient, its torque-relevant contribution scales as $(K_3/M)\sin^5\theta_M$ in the equilibrium condition and is therefore amplified precisely where the switching features are most pronounced. Importantly, the fact that $K_3/M$ can become comparable to, or even exceed, $K_2/M$ over some temperature range is not intrinsically unphysical. This is because these are effective anisotropy coefficients in an angular expansion rather than independent microscopic energy scales and the relative importance of $K_2$ and $K_3$ depends on the instantaneous $\theta_M$ through different powers of $\sin\theta_M$. In other words, a crossing of coefficients does not imply a crossing of physical contributions at all angles. It rather signals that higher-order stiffness near the plane becomes comparable to the quartic stiffness as temperature evolves.

The extracted hierarchy – $K_1 < 0$ (easy-plane tendency), $K_2 > 0$, and $K_3 > 0$ (additional near-plane stiffness) – therefore provides a compact, quantitative description of a multi-axis anisotropy that stabilizes a coherent canted state and controls how the magnetization approaches the plane under field. Additionally, our simulation of the $\theta_c(T)$ with $K_1$-$K_2$-$K_3$ higher-order anisotropy model supports the non-monotonic temperature dependence when Zeeman energy competes with the anisotropy energy scale (see **Fig. 3d** dashed blue-line). We explain the detailed simulation process in Supplementary Information, Section 6.

**Figure 4** provides an interpretation of the anomalous angular transport response reported in Ref. [9]. In **Fig. 4a**, the in-plane resistivity $\rho_{ab}$ plotted as a function of the $\theta_H$ spans the rotational range from $H//c$ to $H//ab$ and exhibits a pronounced, sharply enhanced angular response near the in-plane configuration ($H//ab$), i.e., a narrow angular window in $\theta_H$ where $\rho_{ab}$ changes most rapidly. Such a presentation, however, uses $\theta_H$ as the control variable, whereas the electronic structure and SOC-driven gap modulation are expected to follow the macrospin orientation. To bridge this gap, we use the anisotropy parameters that we obtained, and compute the equilibrium magnetization angle $\theta_M(\theta_H)$ at between 1 and 9 T, which is presented in **Fig. 4b**. The resulting curves show non-linear behavior over the full range, implying that an apparently narrow angular feature in $\theta_H$ can correspond to a substantially broader (and more systematic) dependence when expressed in terms of $\theta_M$. In this sense, **Fig. 4b** already suggests a natural magnetic origin for the apparent compression of the response in $\theta_H$. The applied field rotates the magnetization in a highly non-uniform manner across angles, particularly close to the in-plane limit.

Guided by this mapping, we reparameterize the transport data using $\theta_M$. **Figure 4c** plots $\ln(\rho/\rho_0)$ as a function of $\cos\theta_M$, where $\rho_0 \equiv \rho(\theta_H = 0°)$ is the out-of-plane reference value. Because the calculated $\theta_M(\theta_H)$ includes a weak hysteretic effect, we correct it when converting $\rho(\theta_H)$ to $\rho(\theta_M)$. This ensures that $\ln(\rho/\rho_0)$ reaches its maximum as $\cos\theta_M$ approaches 0, as expected for the in-plane configuration. Then, a notable outcome is that the data follow an approximately linear trend with a near-plane slope that is essentially field-independent to leading order as close to the in-plane limit ($\cos\theta_M \simeq 0$). We quantify this by fitting only the 1 T dataset with a line of slope of -47.9 (red solid-line in **Fig. 4c**), and then applying only vertical offsets for 2, 3, and 5 T to demonstrate the same field-independent slope. This observation connects naturally to the simple transport motif proposed by one of the present authors. If the low-temperature resistivity is dominated by activated transport, $\rho \propto \exp[\Delta(\theta_M)/k_BT]$, then $\ln\left(\frac{\rho}{\rho_0}\right) \approx \frac{\Delta(\theta_M) - \Delta(\theta_M=0°)}{k_BT}$. The near-plane linearity in **Fig. 4c** therefore suggests that, in this angular range, the effective gap can be expanded to leading order as $\Delta(\theta_M) \simeq \Delta_{ab} - A\cos\theta_M$, where $A$ is a prefactor, so that $\ln(\rho/\rho_0)$ becomes approximately proportional to $\cos\theta_M$. For comparison, the inset of **Fig. 4c** presents the same plotted against $\cos\theta_H$. In this representation, the peaky response near the in-plane configuration at 1 and 2 T data, which already evident in **Fig. 4a**, remains prominent and even the 1, 2, and 5 T curves exhibit noticeably different near-plane slopes. This contrast indicates that much of the apparent peakiness and field dependence in $\rho(\theta_H)$ stems from the non-linear mapping from $\theta_H$ to $\theta_M$, and that $\theta_M$ therefore provides a more systematic control variable for organizing the angular transport response.

Note that the pronounced deviation from linearity as $\cos\theta_M \to 1$ (approaching the out-of-plane limit) suggests that additional contributions become relevant away from the near-plane regime, such as higher-order angular dependence of $\Delta(\theta_M)$ and/or angle-dependent prefactors or mixed conduction channels. A detailed microscopic identification of these contributions is beyond the scope of the present work.

**Conclusion**

In summary, torque magnetometry and ESR provide complementary constraints on the anisotropy energy of Mn$_3$Si$_2$Te$_6$. Torque establishes a robust, directly observable switching-angle metric extracted from local extrema in $d(\tau/H)/d\theta_H$, revealing a coherent canted state that persists over a wide field range at low temperature. ESR independently constrains the leading anisotropy fields $K_1/M$ and $K_2/M$. We show that a $K_1$-$K_2$-only free-energy model cannot reproduce the full $\theta_c(H)$ evolution, and that a symmetry-allowed sixth-order term is required to capture the near in-plane stiffness that governing high-field behavior. Using the resulting $K_1$-$K_2$-$K_3$ parameter set, we compute the equilibrium mapping $\theta_M(\theta_H)$ and reparameterize angular transport in terms of $\theta_M$. This replot collapses the near the in-plane limit transport response onto an approximately field-independent linear trend in $\ln(\rho/\rho_0)$ versus $\cos\theta_M$, consistent with a leading-order gap modulation controlled by spin orientation. Hence, our results establish a quantitative magnetic basis for connecting coherent spin canting and higher-order anisotropy to the pronounced angular transport response in this vdW-like ferrimagnetic semiconductor.

## Experimental setup

### A. Single crystal growth

Single crystals of $Mn_3Si_2Te_6$ were synthesized by a high-temperature self-flux method [9]. High-purity Mn (99.95 %), Si (99.999 %), and Te (99.999 %) were mixed in a molar ratio of 1:2:6 and loaded into an alumina crucible. A second empty alumina crucible was placed above the charge, separated by quartz wool, and the crucible assembly was sealed in an evacuated quartz ampoule. The ampoule was heated to 1000 °C over 12 hours and held for 24 hours to homogenize the melt. It was then cooled slowly to 700 °C over 150 hours and maintained at 700 °C for an additional 12 hours. Finally, the ampoule was removed from the furnace and centrifuged to decant the excess flux and collect single crystals. All crystals used in this study (Sample #1 to #4) were selected from the same growth batch to ensure consistent sample quality and growth conditions.

### B. Magnetic torque and magnetization measurement

The used samples in this study were characterized by magnetic torque and SQUID magnetometry. Magnetic torque measurements were performed using piezoresistive cantilevers with controlled field orientation. For Sample#1, measurements were carried out in an Oxford Instruments vector-magnet system, where the field angle was tuned using the vector-field capability without a mechanical rotator. In this configuration, magnetic fields up to 0.8 T were applied. For Sample#2 and #4, torque measurements were performed in a Cryogen Free Measurement System (CFMS) platform and a 9 T Physical Property Measurement System (PPMS, Quantum Design), respectively, both equipped with a mechanical rotator that enabled angular-dependent measurements in magnetic fields up to 9 T. In all setups, the torque signal was recorded as a function of the field angle $\theta_H$ over a wide temperature range (2 – 90 K). The reproducibility of the key torque features across Samples #1, #2 and #4 is presented in the maintext and Supplementary Information, Section 2.

Magnetic susceptibility measurements were carried out using a SQUID magnetometer (MPMS3, Quantum Design), with the magnetic field applied along both in-plane and out-of-plane crystallographic directions.

### C. Electron spin resonance measurement

ESR measurements were performed using a custom-built setup equipped with a tunable microwave source covering the 1–60 GHz frequency range [22]. The Sample#3 was placed in a cryogenic environment under an external magnetic field. Microwave absorption was detected via the bolometric response of a resistive thermometer located beneath the sample.

A low-frequency amplitude modulation of the microwave was employed, and both the amplitude ($\Delta R$) and phase ($\Phi$) of the thermometer signal were recorded using a lock-in amplifier. The ESR signals were obtained by sweeping the magnetic field at fixed microwave frequencies or by varying the frequency under constant field conditions.

### Acknowledgement

The authors are grateful to Chang-Jong Kang and Sanghoon Kim for their helpful discussions. This research was supported by Global-Learning & Academic research institution for Master's·PhD students, Postdocs (G-LAMP) Program of the National Research Foundation of Korea (NRF) grant funded by the Ministry of Education (No. RS-2025-25419722). This work is supported by the National Research Foundation of Korea (NRF) (No. RS-2025-25419722, RS-2025-25453111). J.L. was supported by the National Research Council of Science & Technology (NST) grant by the Korea government (MSIT) (No. GTL24042-000).

### Author Contributions

This work was initiated by J.S.K. and C.W.C.. C.W.C. conducted and coordinated all measurements, interpreted the results. B.K. grew the single crystals and conducted the torque and ESR measurements, with the help of I.C., J.G., J.L., S.-Y.P., J.C., and Y.J.. S.K. and S.P. performed the SQUID magnetization measurements. B.A.P. contributed to the ESR measurements. J.S.K. supervised the project. The manuscript was prepared by C.W.C. and all authors were involved in discussions and contributed to the manuscript.

**Figure 1**

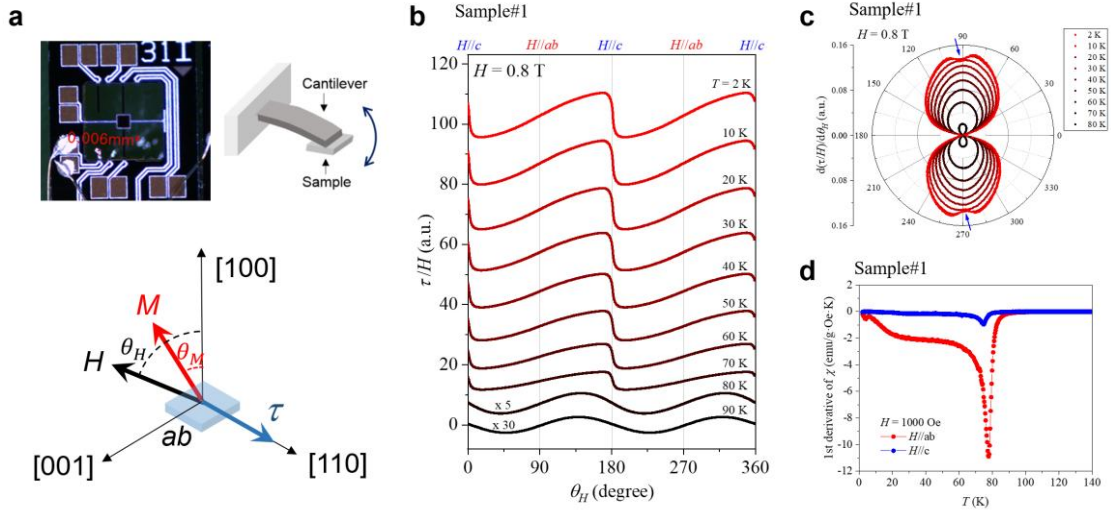

**Figure 1. Torque magnetometry setup and a torque-based signature of coherent spin canting in Mn$_3$Si$_2$Te$_6$. a,** Optical image of a representative single crystal mounted on a piezoresistive cantilever and a schematic of the torque geometry. The magnetic field $H$ is rotated in a plane containing the crystallographic *c*-axis and an in-plane direction. $\theta_H$ and $\theta_M$ denote the field angle measured from the *c*-axis and the corresponding equilibrium magnetization angle, respectively. **b,** Angular dependence of the normalized torque $\tau/H$ measured at $H = 0.8$ T between 2 and 90 K. Vertical offsets are added for clarity. **c,** Polar plots of the angular derivative $d(\tau/H)/d\theta_H$ at $H = 0.8$ T. The low-temperature anomaly (marked by arrows) emerges below 10 K. **d,** Temperature dependence of $d\chi/dT$ for *H*//*ab* and *H*//*c*, yielding $T_c = 78$ K.

**Figure 2**

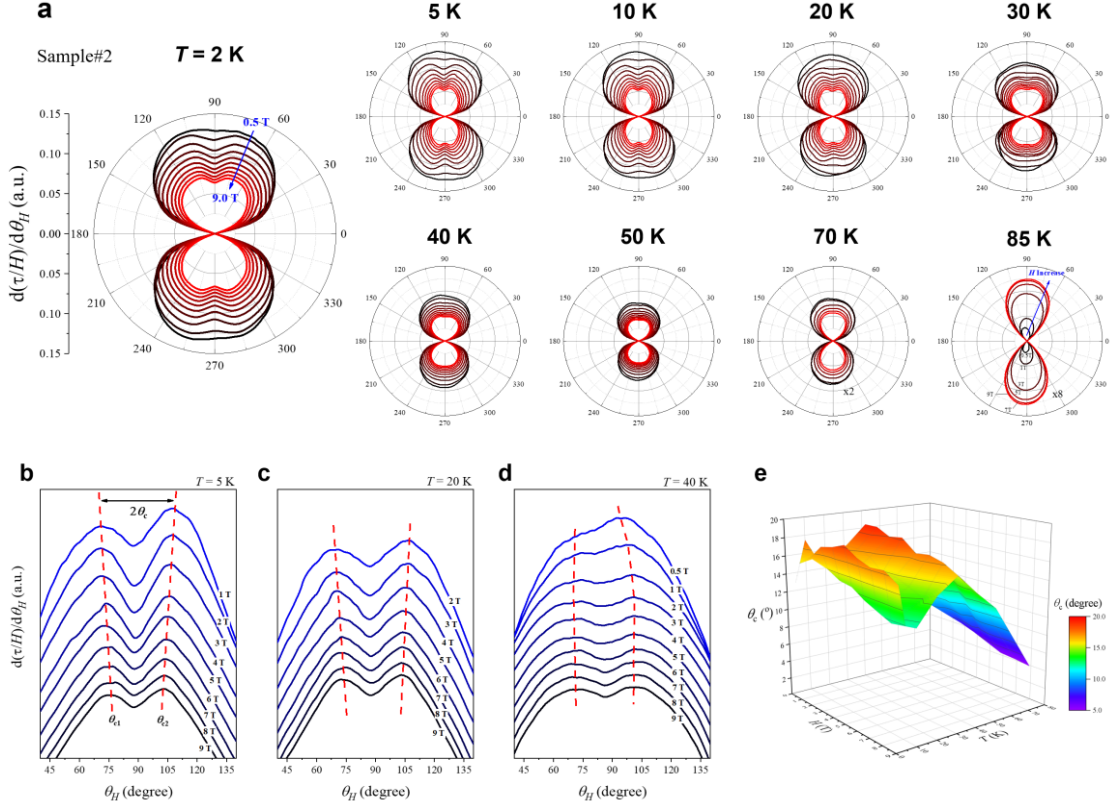

**Figure 2. Field- and temperature evolution of the canting-related anomaly in torque derivative for Mn$_3$Si$_2$Te$_6$ and definition of the switching-angle. a,** Polar plots show the angular derivative $d(\tau/H)/d\theta_H$ measured as a function of $\theta_H$ at selected temperatures from 2 to 85 K. Except for 70 K and 85 K, each panel contains traces taken at $H$ = 0.5 to 9 T (red to black). See arrow for the field-increase direction in representative panels at 2 K. At low temperatures (2 – 10 K), the two-lobe pattern exhibits clear distortions and a systematic shift of the prominent features away from the high-symmetry directions, indicative of a canted equilibrium magnetization. With increasing temperature, the overall amplitude decreases and the anomaly gradually weakens. Near $T_c$ (70 K) the response becomes nearly two-fold symmetric. **b-d,** Representative traces of $d(\tau/H)/d\theta_H$ plotted as a function of field angle $\theta_H$ in the vicinity of the in-plane orientation for selected temperatures: 5 K (b), 20 K (c), and 40 K (d). For each field, two prominent extrema are identified at $\theta_1$ and $\theta_2$ (red dashed lines), and the separation defines the switching-angle metric $2\theta_c \equiv \theta_2 - \theta_1$. Curves are vertically offset for clarity. **e,** 3D color map of $\theta_c(T,H)$ obtained from the extrema positions for all the torque data, visualizing the persistence and systematic evolution of the switching markers across temperature and magnetic field. The color scale indicates the magnitude of $\theta_c$.

**Figure 3**

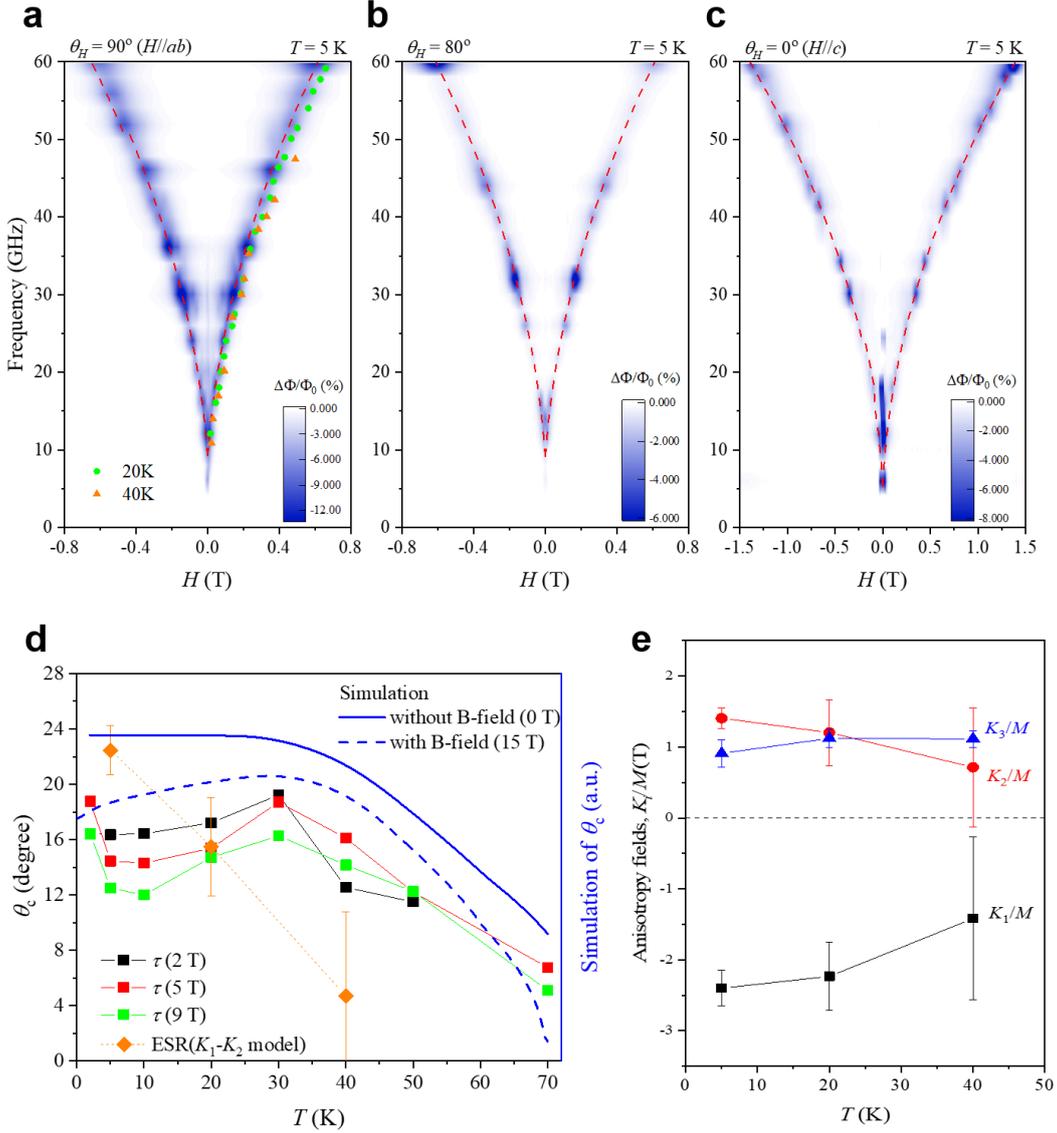

**Figure 3. ESR constraints and the emergence of a sixth-order anisotropy term. a-c,** Field-frequency maps of the ESR phase response $\Delta\Phi/\Phi_0$ measured at $T = 5$ K for three field orientations: $\theta_H = 90°$ (H//ab) (a), $\theta_H = 80°$ (b), and $\theta_H = 0°$ (H//c) (c). Color scale denotes $\Delta\Phi/\Phi_0$. Red dashed lines indicate fitted resonance dispersions obtained from a global analysis based on the anisotropy-energy model (see Supplementary Information, Section 4). In panel (a), symbols overlay additional resonance points measured at 20 K (green circles) and 40 K (orange triangles) for H//ab. **d,** Temperature dependence of the torque-derived switching-angle $\theta_c$ at representative fields. For comparison, the orange diamonds show $\theta_c$ predicted from the ESR-only fit using the $K_1$-$K_2$ model, demonstrating that the ESR-constrained $K_1$-$K_2$ terms alone cannot reproduce the persistence of $\theta_c$ at high fields. Blue curves show simulations of $\theta_c(T)$ including a sixth-order contribution, highlighting the role of near in-plane stiffness (solid line: without an additional field-induced contribution; dashed line: with an additional field-induced contribution). **e,** Extracted anisotropy fields $K_1/M$, $K_2/M$, and $K_3/M$ as a function of temperature. $K_1/M$ and $K_2/M$ are constrained primarily by ESR, while $K_3/M$ is determined by matching the high-field $\theta_c(H,T)$ behavior from torque (error bars reflect fitting uncertainty).

**Figure 4**

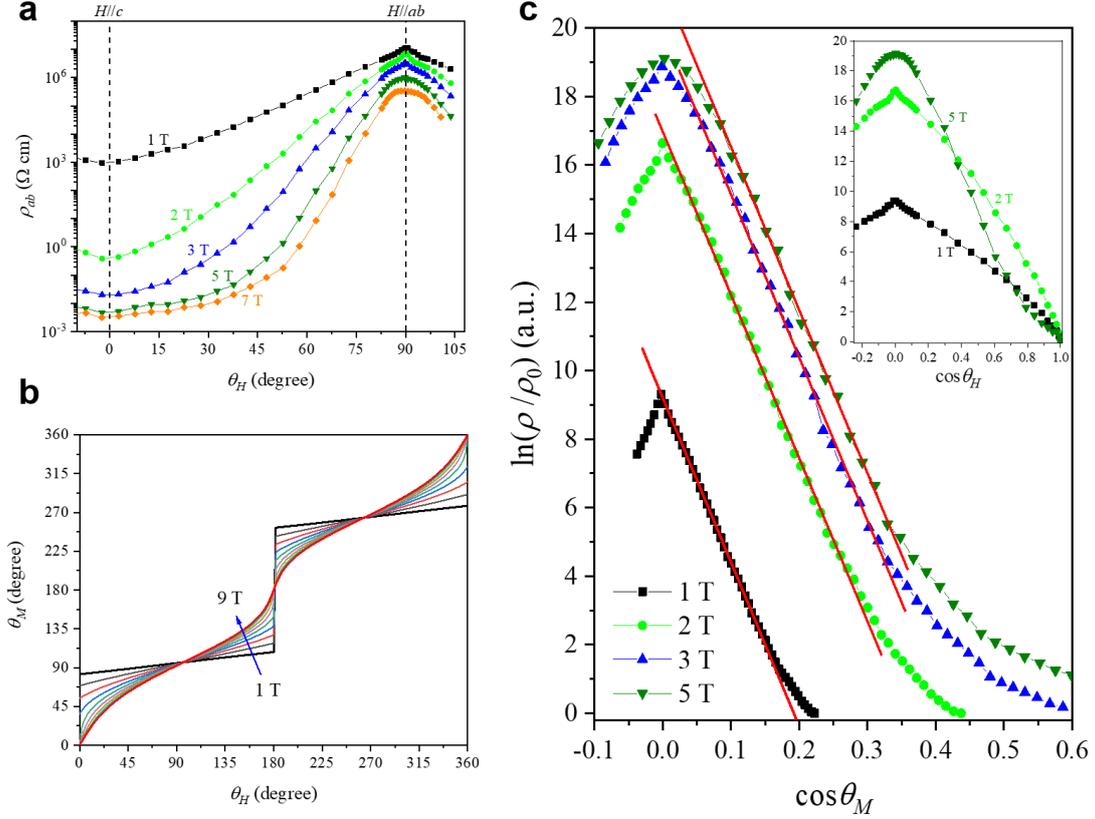

**Figure 4. Reparameterizing CAMR in terms of the equilibrium magnetization angle. a,** Colossal angular magnetoresistance reported by *Seo et al*. [9]. In-plane resistivity $\rho_{ab}$ is plotted as a function of the applied-field angle $\theta_H$ for several magnetic fields. Vertical dashed lines mark *H//c* ($\theta_H = 0°$) and *H//ab* ($\theta_H = 90°$), respectively. **b,** Calculated equilibrium magnetization angle $\theta_M(\theta_H)$ obtained using the anisotropy parameters extracted in this work. Curves are shown for representative fields from 1 to 9 T, highlighting a strongly non-linear behavior between the field direction and the magnetization orientation. **c,** Transport data from panel (a) re-expressed in terms of $\theta_M$ using the relation in panel (b): $\ln(\rho/\rho_0)$ plotted versus $\cos\theta_M$, where $\rho_0 \equiv \rho(\theta_H = 0°)$ is the out-of-plane reference resistivity. The red solid-line shows a linear fit to the 1 T dataset (slope -47.9). The same slope is vertically shifted for 2, 3, and 5 T to emphasize a field-independent near-plane slope. **Inset**, $\ln(\rho/\rho_0)$ plotted against $\cos\theta_H$ for selected fields (1, 2, and 5 T), illustrating that the near-plane response remains strongly field-dependent when expressed in terms of $\theta_H$.